\documentstyle[aps,prd,epsf]{revtex}
\input{psfig}

\begin{document}
\draft
\title{
Double Diffraction Dissociation 
at the Fermilab Tevatron Collider
}
\maketitle
\vglue 1em
\centerline{\large The CDF Collaboration}
\vglue 1em
\centerline{\large (Submitted to Physical Review Letters)}
\vglue 1em

\font\eightit=cmti8
\def\r#1{\ignorespaces $^{#1}$}
\hfilneg
\begin{sloppypar}
\noindent
T.~Affolder,\r {23} H.~Akimoto,\r {45}
A.~Akopian,\r {37} M.~G.~Albrow,\r {11} P.~Amaral,\r 8  
D.~Amidei,\r {25} K.~Anikeev,\r {24} J.~Antos,\r 1 
G.~Apollinari,\r {11} T.~Arisawa,\r {45}
% ...
\end{sloppypar}
\vskip .026in

\vskip .026in
\begin{center}
\r 1  {\eightit Institute of Physics, Academia Sinica, Taipei, Taiwan 11529, 
Republic of China} \\
\r 2  {\eightit Argonne National Laboratory, Argonne, Illinois 60439} \\
\r 3  {\eightit Istituto Nazionale di Fisica Nucleare, University of Bologna,
I-40127 Bologna, Italy} \\
\r 4  {\eightit Brandeis University, Waltham, Massachusetts 02254} \\
\r 5  {\eightit University of California at Davis, Davis, California  95616} \\
\r 6  {\eightit University of California at Los Angeles, Los 
Angeles, California  90024} \\  
\r 7  {\eightit Instituto de Fisica de Cantabria, CSIC-University of Cantabria, 
39005 Santander, Spain} \\
\r 8  {\eightit Enrico Fermi Institute, University of Chicago, Chicago, 
Illinois 60637} \\
\r 9  {\eightit Joint Institute for Nuclear Research, RU-141980 Dubna, Russia}
\\
\r {10} {\eightit Duke University, Durham, North Carolina  27708} \\
\r {11} {\eightit Fermi National Accelerator Laboratory, Batavia, Illinois 
60510} \\
\r {12} {\eightit University of Florida, Gainesville, Florida  32611} \\
\r {13} {\eightit Laboratori Nazionali di Frascati, Istituto Nazionale di Fisica
               Nucleare, I-00044 Frascati, Italy} \\
\r {14} {\eightit University of Geneva, CH-1211 Geneva 4, Switzerland} \\
\r {15} {\eightit Glasgow University, Glasgow G12 8QQ, United Kingdom}\\
\r {16} {\eightit Harvard University, Cambridge, Massachusetts 02138} \\
\r {17} {\eightit Hiroshima University, Higashi-Hiroshima 724, Japan} \\
\r {18} {\eightit University of Illinois, Urbana, Illinois 61801} \\
\r {19} {\eightit The Johns Hopkins University, Baltimore, Maryland 21218} \\
\r {20} {\eightit Institut f\"{u}r Experimentelle Kernphysik, 
Universit\"{a}t Karlsruhe, 76128 Karlsruhe, Germany} \\
\r {21} {\eightit Center for High Energy Physics: Kyungpook National
University, Taegu 702-701; Seoul National University, Seoul 151-742; and
SungKyunKwan University, Suwon 440-746; Korea} \\
\r {22} {\eightit High Energy Accelerator Research Organization (KEK), Tsukuba, 
Ibaraki 305, Japan} \\
\r {23} {\eightit Ernest Orlando Lawrence Berkeley National Laboratory, 
Berkeley, California 94720} \\
\r {24} {\eightit Massachusetts Institute of Technology, Cambridge,
Massachusetts  02139} \\   
\r {25} {\eightit University of Michigan, Ann Arbor, Michigan 48109} \\
\r {26} {\eightit Michigan State University, East Lansing, Michigan  48824} \\
\r {27} {\eightit University of New Mexico, Albuquerque, New Mexico 87131} \\
\r {28} {\eightit The Ohio State University, Columbus, Ohio  43210} \\
\r {29} {\eightit Osaka City University, Osaka 588, Japan} \\
\r {30} {\eightit University of Oxford, Oxford OX1 3RH, United Kingdom} \\
\r {31} {\eightit Universita di Padova, Istituto Nazionale di Fisica 
          Nucleare, Sezione di Padova, I-35131 Padova, Italy} \\
\r {32} {\eightit University of Pennsylvania, Philadelphia, 
        Pennsylvania 19104} \\   
\r {33} {\eightit Istituto Nazionale di Fisica Nucleare, University and Scuola
               Normale Superiore of Pisa, I-56100 Pisa, Italy} \\
\r {34} {\eightit University of Pittsburgh, Pittsburgh, Pennsylvania 15260} \\
\r {35} {\eightit Purdue University, West Lafayette, Indiana 47907} \\
\r {36} {\eightit University of Rochester, Rochester, New York 14627} \\
\r {37} {\eightit Rockefeller University, New York, New York 10021} \\
\r {38} {\eightit Rutgers University, Piscataway, New Jersey 08855} \\
\r {39} {\eightit Texas A\&M University, College Station, Texas 77843} \\
\r {40} {\eightit Texas Tech University, Lubbock, Texas 79409} \\
\r {41} {\eightit Institute of Particle Physics, University of Toronto, Toronto
M5S 1A7, Canada} \\
\r {42} {\eightit Istituto Nazionale di Fisica Nucleare, University of Trieste/
Udine, Italy} \\
\r {43} {\eightit University of Tsukuba, Tsukuba, Ibaraki 305, Japan} \\
\r {44} {\eightit Tufts University, Medford, Massachusetts 02155} \\
\r {45} {\eightit Waseda University, Tokyo 169, Japan} \\
\r {46} {\eightit University of Wisconsin, Madison, Wisconsin 53706} \\
\r {47} {\eightit Yale University, New Haven, Connecticut 06520} \\
\r {(\ast)} {\eightit Now at Carnegie Mellon University, Pittsburgh,
Pennsylvania  15213} \\
\r {(\ast\ast)} {\eightit Now at Northwestern University, Evanston, Illinois 
60208}
\end{center}
\vglue 1em
\centerline{\large Abstract}

\begin{abstract}
We present results from a measurement of double diffraction 
dissociation in $\bar{p}p$
collisions at the Fermilab Tevatron collider.
The production cross section for events 
with a central pseudorapidity gap of 
width $\Delta\eta^0>3$ (overlapping $\eta=0$) is found to be 
$4.43\pm 0.02\mbox{(stat)}{\pm 1.18}\mbox{(syst) mb}$
[$3.42\pm 0.01\mbox{(stat)}{\pm 1.09}\mbox{(syst) mb}$]
at $\sqrt{s}=1800$ [630] GeV.
Our results are compared with previous measurements
and with predictions based on Regge
theory and factorization.
\end{abstract}

\pacs{PACS number(s): 13.85.Ni}

Double diffraction (DD) dissociation 
is the process in which two colliding hadrons 
dissociate into clusters of particles producing
events with a central pseudorapidity~\cite{pseudo} gap 
(region of pseudorapidity devoid of particles), 
as shown in  Fig.~\ref{f:pictdd}. 
This process is similar to single diffraction (SD) dissociation,
in which one of the incident hadrons dissociates
while the other escapes as a leading (highest momentum) particle.
Events with pseudorapidity gaps are presumed to be due to the exchange 
across the gap of a Pomeron~\cite{Regge}, which in QCD is a color 
singlet state with vacuum quantum numbers. 

Previous measurements of DD have been performed only 
over limited pseudorapidity regions for 
$\bar{p}p$ collisions at $\sqrt{s}=200$ and 900 GeV~\cite{UA5dd},
for exclusive and semi-inclusive dissociation channels 
at lower energies~\cite{ISR2,ISR}, 
e.g $pp\rightarrow (p\pi^+\pi^-)(p\pi^+\pi^-)$ 
or $pp\rightarrow (p\pi^+\pi^-)+X$, and for $\gamma p$ interactions at  
the DESY $ep$ collider HERA~\cite{h1dd}. 
The present measurement, based on a study of
central rapidity gaps in minimum bias events from 
$\bar pp$ collisions at $\sqrt{s}=1800$ and 630 GeV collected by the 
Collider Detector at Fermilab (CDF), covers a wide $\eta$ range,
allowing comparisons with theoretical predictions on both 
$\eta$-dependence and normalization.

To facilitate our discussion, we begin by defining the relevant 
variables~\cite{dinophysrep}.
We use $s$ and  $t$ for the square of the c.m.s.  
energy and 4-momentum transfer between the two incident hadrons, 
$\xi$ for the fractional momentum loss of the leading hadron in SD,
and $\eta$ for pseudorapidity.
For $\bar pp$ double diffraction dissociation into masses $M_1$ and $M_2$,
we define the {\em nominal} pseudorapidity gap as 
$\Delta\eta\equiv \ln\frac{\textstyle{s}\textstyle{s_{\circ}}}{M_1^2M_2^2}$, 
where ${\textstyle{s_{\circ}}}\equiv 1$ GeV$^2$; on average, the nominal gap 
is approximately equal to the true rapidity gap in an event.  
A variable defined as $s'\equiv M_1^2M_2^2/{\textstyle{s_{\circ}}}$ 
can be thought of as the generalization of $s'=M^2$ for SD, 
since in both cases $\ln \frac{s'}{s_{\circ}}$ represents 
the pseudorapidity region accessible to the dissociation products 
of the diffractive sub-system(s). For $\bar pp$ SD with 
$M_2=m_p\approx 1\;{\rm GeV}$, $s'=M_1^2$ and 
$\xi=e^{-\Delta\eta}={\textstyle s'}/{\textstyle s}$. 
 
Diffraction has traditionally been treated theoretically in the framework 
of Regge phenomenology~\cite{Regge}.
At large $\Delta\eta$, where Pomeron exchange is dominant~\cite{dinophysrep},
the SD cross section is given by the triple-Pomeron amplitude, 
\begin{equation}{d^{2}\sigma_{SD}\over dtd\Delta \eta}= 
  \left[{\beta^{2}(t)\over 16\pi} e^{2[\alpha(t)-1]\Delta \eta}\right]
  \left[\kappa\beta^{2}(0)
{\left(\frac{s'}{\textstyle s_{\circ}}\right)}^{\alpha(0)-1}\right] 
\label{eq:sd} 
\end{equation}
where $\alpha(t)$ 
is the Pomeron trajectory, $\beta(t)$ the coupling of the 
Pomeron to the (anti)proton, and $\kappa\equiv g(t)/\beta(0)$ the ratio 
of the triple-Pomeron to the Pomeron-proton  
couplings; 
we use $\alpha(t)=\alpha(0)+\alpha't=1.104+0.25t$~\cite{CMG}, 
$\beta(0)=4.1$ mb$^{1/2}$~\cite{CMG}, 
and $g(t)=0.69$ mb$^{1/2}$ ($\Rightarrow \kappa=0.17$)~\cite{dino_montanha}.
The second factor  of Eq.~\ref{eq:sd} 
has the form of the Pomeron-proton total cross section at 
the sub-energy $\sqrt{s'}$, while the first factor 
can be thought of as a rapidity gap probability~\cite{dino_gap}.  
Measurements on SD 
have shown that Eq.~\ref{eq:sd}, which is based on Regge factorization, 
correctly predicts the $\Delta\eta$ dependence for $\Delta\eta>3$, 
but fails to predict the 
energy dependence of the overall normalization, which 
at $\sqrt s=1800$ GeV is found to be suppressed by an order of 
magnitude~\cite{CDF_SD,dino_flux}. 
It is generally believed that this breakdown of factorization is imposed by 
unitarity constraints~\cite{GLM}. 
Phenomenologically, it has been shown that
normalizing the integral of the gap probability
(first factor in Eq.~1)  over all phase space to unity yields the correct energy
dependence~\cite{dino_montanha,dino_flux}.

Using factorization, the DD differential 
cross section may be expressed 
in terms of the SD and elastic scattering cross sections as~\cite{dinophysrep}
\begin{eqnarray}{d^{3}\sigma_{DD}\over dtdM_1^2dM_2^2} & = & 
  {d^{2}\sigma_{SD}\over dtdM_1^2}{d^{2}\sigma_{SD}\over dtdM_2^2}
  \,/\,{d\sigma_{el}\over dt} 
\nonumber\\
   & = & {[\kappa \beta_1(0)\beta_2(0)]^2\over 16\pi}\,
  {{s^{2[\alpha(0)-1]}\, e^{b_{DD}t}\over (M_1^2M_2^2)^{1+2[\alpha(0)-1]}}}
\end{eqnarray}
where $b_{DD}=2\alpha'\ln{({\textstyle s}{\textstyle s_{\circ}}/M_1^2M_2^2)}$.
Changing variables from $M_1$ and $M_2$ to $\Delta\eta$ and 
$\eta_c=\ln\frac{M_2}{M_1}$, where $\eta_c$ is the center of the 
rapidity gap, yields (setting $\beta_1=\beta_2\Rightarrow\beta$)  
\begin{equation}{d^{3}\sigma_{DD}\over dtd\Delta \eta d\eta_c}=
\left[{\kappa\beta^{2}(0)\over 16\pi} e^{2[\alpha(t)-1]\Delta \eta}\right]
\left[\kappa\beta^{2}(0){{\left(\frac{s'}{\textstyle s_{\circ}}\right)}}^
{\alpha(0)-1}
\right] 
\label{eq:dd} 
\end{equation}
This expression is strikingly similar to Eq. \ref{eq:sd},
except that, 
since the gap is now not adjacent to  a leading (anti)proton, 
$\eta_c$ is treated as an independent variable.
The question that arises naturally is whether Eq.~\ref{eq:dd} correctly 
predicts the differential DD cross section 
apart from an overall normalization factor, 
as is the case with Eq.~\ref{eq:sd} for SD.
The answer to this question, and the suppression in overall normalization 
relative to that observed in SD, could provide a crucial check
on models proposed to account for the factorization breakdown
observed in SD.

The components of CDF~\cite{detector} relevant to this study 
are the central tracking chamber (CTC), the calorimeters, 
and two scintillation beam-beam counter (BBC) 
arrays.
The CTC tracking efficiency varies from $\sim60$\% for $p_T=300$ 
MeV to over 95\% for $p_T>400$ MeV within $|\eta|<1.2$, and 
falls monotonically beyond 
$|\eta|=1.2$ approaching zero at $|\eta|\sim 1.8$.  
The calorimeters have projective tower geometry and cover the 
regions $|\eta|<1.1$ (central), $1.1<|\eta|<2.4$ (plug), and $2.2<|\eta|<4.2$ 
(forward). The $\Delta \eta \times \Delta \phi$ tower dimensions are 
$0.1\times 15^{\circ}$ for the central and $0.1\times 5^{\circ}$ for the 
plug and forward calorimeters. The BBC arrays cover the region $3.2<|\eta|<5.9$.
 
Events collected by triggering on a BBC coincidence between the proton 
and antiproton sides of the detector 
comprise the CDF {\em minimum-bias} (MB) data
sample. Two MB data sets are used in this analysis, 
consisting of $1.0\times 10^6$ ( $1.6\times 10^6$) events 
at $\sqrt{s}=1800$  [630] GeV
obtained at average instantaneous luminosities of 
$2.5\times 10^{30}$ ($9.6\times 10^{29}$) 
cm$^{-2}$sec$^{-1}$.
At these luminosities, the fraction of {\em overlap} 
events due to multiple interactions is estimated to be 20.7 (6.5)\%. 
To reject overlap events, we accept only events with no more than one 
reconstructed vertex within $\pm 60$ cm from the center of the detector.

The method we use to search for a DD signal is based on the  approximately
flat dependence of the event rate on $\Delta\eta$ expected for DD events,
as seen by setting $\alpha(t)=1.104+0.25t$ in Eq.~\ref{eq:dd}, 
compared to the exponential dependence expected for
non-diffractive (ND) events where rapidity gaps are due to random multiplicity
fluctuations. Thus, in a plot of event rate versus $\Delta\eta$, the DD signal
will appear as the flattening at large $\Delta\eta$ of an exponentially
falling distribution. For practical considerations, our analysis is based
on {\em experimental} gaps defined as
$\Delta\eta_{exp}^0\equiv\eta_{max}-\eta_{min}$,
where ($\eta_{min}$) $\eta_{max}$ is the $\eta$ of the ``particle'' closest to
$\eta=0$ in the (anti)proton direction (see Fig.~1).  
A ``particle" is a reconstructed
track in the CTC, a calorimeter tower with energy above a given threshold,
or a BBC hit. The (uncorrected) tower energy thresholds used, chosen to lie
comfortably above noise level, are $E_T=0.2$ GeV for the central and plug and
$E=1$ GeV for the forward calorimeters. At the calorimeter interfaces near
$|\eta|\sim 0$, 1.1  and $\sim 2.4$, where the noise level is higher,
$|\eta|$-dependent thresholds are used. The DD signal is extracted by fitting
the measured $\Delta\eta_{exp}^0$ distribution with expectations
based on a Monte Carlo (MC) simulation incorporating SD, DD and ND
contributions.  The same thresholds are used in the MC simulations after
dividing the generated particle energy by an $\eta$-dependent energy
calibration coefficient representing the ratio of true to measured
(uncorrected) calorimeter energy~\cite{suren}. For charged-particle tracks,
the MC generation is followed by a detector simulation.

Figure~\ref{f:etamax2d} shows lego histograms of events versus $\eta_{max}$
and $-\eta_{min}$ for data and for Monte Carlo generated ND, SD and
DD events at $\sqrt{s}=1800$ GeV. A uniform $\eta$-distribution was assumed
for particles within a calorimeter tower. The observed structure in the
distributions along $\eta_{max(min)}$ is caused by the variation of the tower
energy threshold with $|\eta|$. The bins at $|\eta_{max(min)}|=3.3$ contain
all events within the BBC range of $3.2<|\eta_{max(min)}|<5.9$.

The diffractive Monte Carlo program is a modified version of that used in
Ref.~\cite{CDFsigmatot}, incorporating the differential cross sections 
of Eqs.~\ref{eq:sd} and ~\ref{eq:dd}. 
Non-diffractive interactions are simulated using PYTHIA~\cite{PYTHIA}.
The data distribution in Fig.~\ref{f:etamax2d} has a larger fraction 
of events at large $|\eta_{max(min)}|$ than either the ND or the SD 
Monte Carlo generated distributions.
From the previously measured SD cross section~\cite{CDF_SD} 
and the MC determined 
fraction of SD events triggering both BBC arrays,
the fraction of SD events in our 1800 [630] GeV 
data sample  is estimated to be 2.7\% [2.4\%]. 
A combination of 97.3\% ND and 2.7\% SD generated events cannot 
account for the data at large $|\eta_{max(min)}|$ in Fig.~\ref{f:etamax2d}.
The simulated DD distribution is approximately flat in $|\eta_{max(min)}|$
and describes the data well when combined with the ND and SD distributions, 
as shown below.  

Figure~\ref{f:gapwidth} presents the number of events 
as a function of $\Delta\eta^0_{exp}$ for 
the 1800 GeV data (points) and for a fit to the data using
a mixture of MC generated DD and ``non-DD" 
(ND plus SD) contributions (solid histogram).
The dashed histogram shows the non-DD contribution.
The agreement between data and MC indicates that, 
as in SD, the shape of the differential DD cross section is 
correctly described by Regge theory and factorization.

At $\sqrt s=1800$ [630] GeV, the fraction of events with 
$\Delta\eta^0_{exp}>3$ (gap fraction)
is $(9.08\pm 0.03)\%$ [$(11.43\pm 0.03)\%$], in which 
the fraction of background 
non-DD events, estimated using the MC simulation, 
is $(23.6\pm 0.6)\%$ [$(29.7\pm 0.6)\%$].
After background subtraction, the DD gap fraction becomes $(6.94\pm 0.06)\%$ 
[$(8.03\pm 0.08)\%$]. The quoted errors are statistical. 
The amount of ND background in the region $\Delta\eta^0_{exp}>3$ 
depends on the tower energy calibration coefficients and thereby on the 
calorimeter tower energy thresholds used in the MC. 
Increasing these thresholds has the effect of decreasing 
the multiplicity in the MC generated events, resulting in larger rapidity gaps 
and hence larger ND backgrounds in the region of $\Delta\eta^0_{exp}$.  
The systematic uncertainty in the background 
is estimated by raising (lowering) the tower thresholds in the MC 
by a factor of 1.25 and 
refitting the data. This change in thresholds increases (decreases)  
the background by a factor of 1.54 (0.52) [1.56 (0.56)]. 

The vertex cut employed to reject events due to 
multiple interactions also rejects single interaction events with extra (fake) 
vertices resulting from track reconstruction ambiguities. 
By comparing the fraction of events surviving the vertex cut with the 
fraction of single interaction events expected from the BBC cross section and 
the instantaneous luminosity, the vertex cut efficiency (fraction of 
single interaction events retained) is found to be $0.87\pm 0.02\mbox{(syst)}$
[$0.90\pm 0.02\mbox{(syst)}$].
In determining the DD gap fraction 
this efficiency is applied only to the total number of events,
since the gap events have low central multiplicities and therefore 
are not likely to have fake vertices.

The measured DD gap fractions, 
which are based on our experimental gap 
definition, $\Delta\eta^0_{exp}\equiv \eta_{max}-\eta_{min}$,
depend on the particle $E_T$ thresholds used.
The correction factors needed to 
transform the measured gap fractions 
to gap fractions corresponding to the gap definition on which Eq.~\ref{eq:dd}
is based, namely $\Delta\eta^0
\equiv \ln\frac{ss_{\circ}}{M_1^2M_2^2}$
($\ln\frac{M_i^2}{\sqrt {ss_{\circ}}}<0,\; i=1,2$), were 
evaluated using the DD Monte Carlo simulation and 
found to be 0.81 [0.75] for $\sqrt s=1800$ [630] GeV.
Correcting the measured DD gap fractions by these factors and for 
the vertex cut efficiency, and 
normalizing the results to our previously measured cross sections of 
$51.2\pm 1.7$~mb [$39.9\pm 1.2$~mb]
for events triggering the BBC arrays,
we obtain $2.51\pm 0.01({\rm stat})\pm 0.08({\rm norm})
{\pm 0.58}{\rm (bg)}$~mb
[$2.16\pm 0.01({\rm stat})\pm 0.06({\rm norm})
{\pm 0.65}{\rm (bg)}$~mb] 
for the DD cross section 
in the region $\Delta\eta^0>3$. 

The trigger acceptance, evaluated from the DD MC simulation, 
is $0.57\pm 0.07({\rm syst})$ [$0.63\pm 0.07)({\rm syst})$].
The uncertainty was estimated by 
considering variations in the simulation of small mass 
diffraction dissociation. The acceptance corrected DD cross sections for 
$\Delta\eta^0>3$ are
$4.43\pm 0.02{\rm (stat)}{\pm 1.18}{\rm (syst)}$ mb 
[$3.42\pm 0.01{(\rm stat)}{\pm 1.09}{\rm(syst)}\;{\rm mb}$].

The corresponding cross sections predicted by Eq.~\ref{eq:dd},
determined by the DD MC simulation, are 
$49.4\pm 11.1\;({\rm syst})$~mb
[$27.7\pm 5.5\;({\rm syst})$~mb], where the uncertainty is due to an 
assigned 10\% systematic error in the triple-Pomeron coupling 
$g(0)=\kappa\beta(0)$~\cite{dino_montanha}.
The ratio (discrepancy factor) of measured to predicted cross sections is 
$D_{DD}=0.09\pm 0.03$ 
[$0.12\pm 0.03$], where the errors include the statistical and all systematic 
uncertainties. 
Recalling that Eq.~\ref{eq:dd} correctly describes the shape of the 
$\Delta\eta^0$ distribution, 
the deviation of $D$ from unity represents a breakdown of factorization
affecting only the overall normalization. This result is 
similar to that observed in SD~\cite{dino_montanha,dino_flux},
where the corresponding discrepancy factors, 
calculated from the fit parameters 
in Ref.~\cite{dino_montanha}, are
$D_{SD}=0.11\pm 0.01$ [$0.17\pm 0.02$].

Our data are compared with the UA5 results~\cite{UA5dd}
in Fig.~\ref{f:ddcomp}. The comparison is made for cross sections 
integrated over $t$ and over all gaps of
$\Delta\eta>3$, corresponding to $\xi=e^{-\Delta\eta}=0.05$
in SD. The extrapolation of our data from $\Delta\eta^0>3$
(gaps overlapping $\eta=0$) to $\Delta\eta>3$ (all gaps)
was made using Eq.~\ref{eq:dd}
and amounts to multiplying the $\Delta\eta^0>3$ cross sections by a 
factor of 1.43 (1.34) at $\sqrt s=1800$ [630] GeV, yielding
  \begin{eqnarray} \lefteqn{\sigma_{DD}(\sqrt{s}=1800\;[630]\;{\rm GeV}, 
\label{sigmadd_2}
  \Delta\eta> 3) =} \qquad \\
\nonumber
  & & 6.32\pm 0.03{\rm (stat)}{\pm 1.7}{\rm (syst)}\;{\rm mb}\qquad \\
\nonumber
  & & [4.58\pm 0.02{(\rm stat)}{\pm 1.5}{\rm(syst)}\;{\rm mb}]
\end{eqnarray}
The reported UA5 cross 
section values were obtained by extrapolating cross sections 
measured over limited large-gap regions to $\Delta\eta>3$ 
using a Monte Carlo simulation in which the $p$ and $\bar p$ dissociated
independently with a $(1/M^2)e^{7t}$ distribution~\cite{UA5sim}. 
For a meaningful 
comparison, we corrected the reported UA5 values by backtracking
to the measured limited $\Delta\eta$ regions using a $(1/M^2)e^{7t}$
dependence and then extrapolating to $\Delta\eta>3$ using Eq.~\ref{eq:dd}.
This correction increases the cross sections by a factor of 1.43 [1.19]
at $\sqrt s=200$ [900] GeV. 
The solid curve in Fig.~\ref{f:ddcomp} was calculated using Eq.~\ref{eq:dd}.
The disagreement between this curve and the data represents the
breakdown of factorization discussed above. The dashed curve represents the 
prediction of the renormalized gap probability model~\cite{dino_flux,dino_gap},
in which the 
integral of the gap probability (first factor in Eq.~\ref{eq:dd} 
over all available phase space)
is normalized to unity. 
The error bands around the curves are due 
to the 10\% uncertainty in the triple-Pomeron coupling~\cite{dino_montanha}.
Within the quoted uncertainties, the data are in agreement with the 
renormalized gap model.

In conclusion, we have measured double diffraction differential 
cross sections in $\bar pp$ collisions at $\sqrt s=$1800 and 630 GeV 
and compared our results with data at $\sqrt s$=200 and 900 GeV and 
with predictions based on Regge theory and factorization. We find 
a factorization breakdown similar in magnitude to that observed in single 
diffraction dissociation. The data are in agreement with the 
renormalized gap probability model~\cite{dino_gap}.

     We thank the Fermilab staff and the technical staffs of the
participating institutions for their vital contributions.  This work was
supported by the U.S. Department of Energy and National Science Foundation;
the Italian Istituto Nazionale di Fisica Nucleare; the Ministry of Education,
Science, Sports and Culture of Japan; the Natural Sciences and Engineering
Research Council of Canada; the National Science Council of the Republic of
China; the Swiss National Science Foundation; the A. P. Sloan Foundation; the
Bundesministerium fuer Bildung und Forschung, Germany; the Korea Science
and Engineering Foundation;
the Max Kade Foundation; and the Ministry of Education, Science and Research
of the Federal State Nordrhein-Westfalen of Germany.

\newpage
\begin{figure}
\vspace*{2in}
\centerline{\psfig{figure=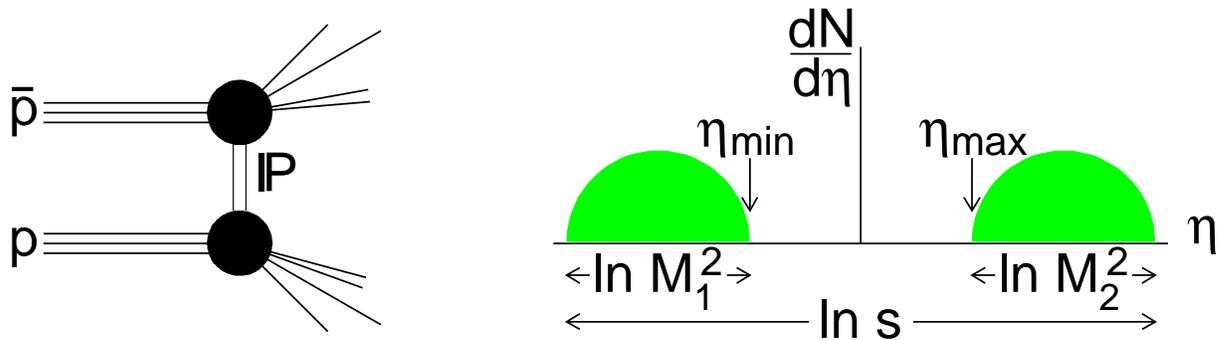,width=0.97\linewidth}}
\vglue -3.65in
\caption{Schematic diagram and event topology of a 
double diffractive interaction, in which a Pomeron ($I\!\!P$) is 
exchanged in a $\bar pp$ collision at center-of-mass energy 
$\protect\sqrt s$ 
producing diffractive masses $M_1$ and $M_2$ separated by a rapidity gap
of width $\Delta\eta=\eta_{max}-\eta_{min}$. The shaded areas represent 
regions of particle production (mass and energy units are in GeV).
}
\end{figure}

\begin{figure}
\centerline{\psfig{figure=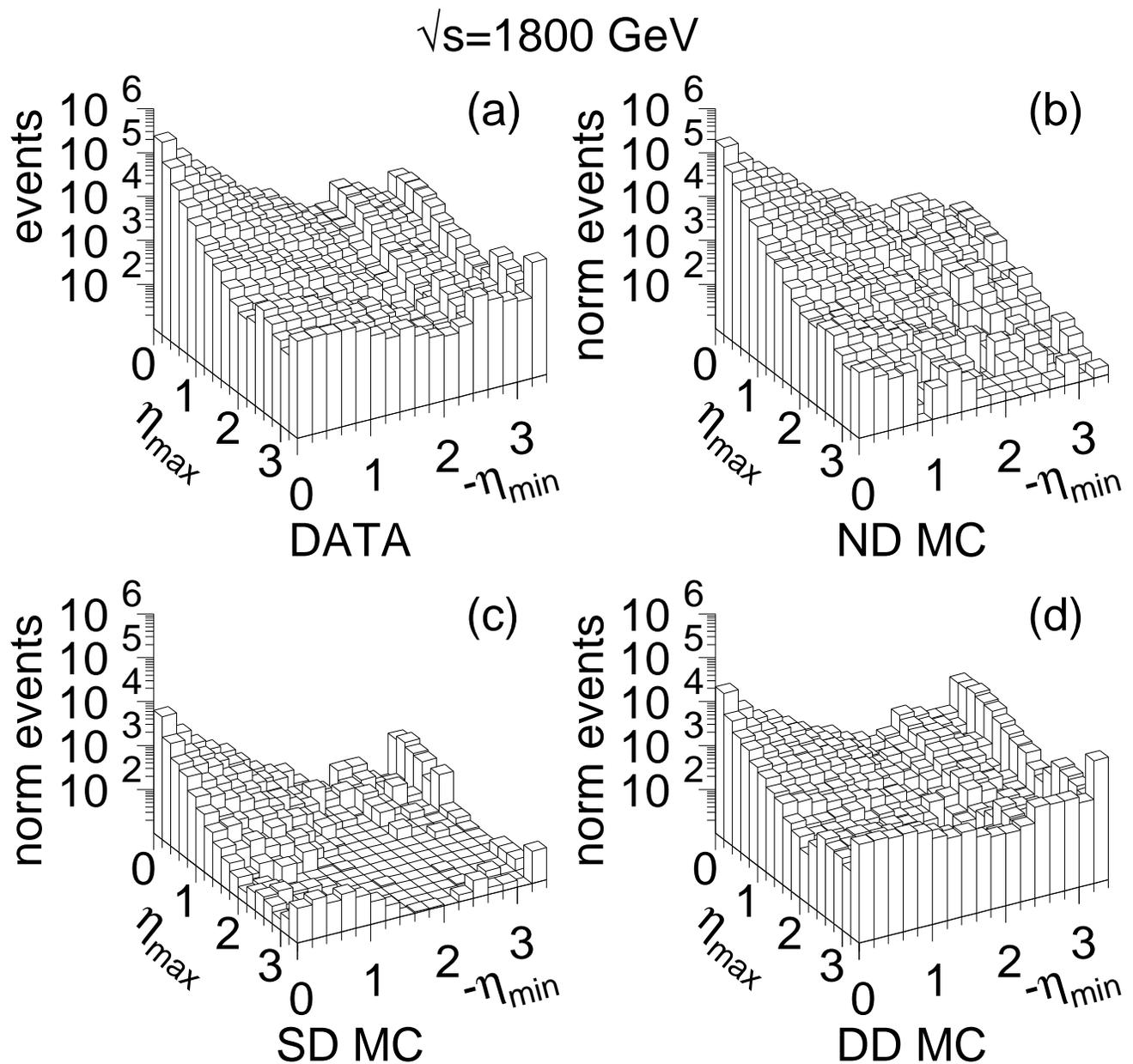,width=0.97\linewidth}}
\vfill
\caption{The number of events as a function of \protect$\eta_{max}$ and 
\protect$-\eta_{min}$,
the $\eta$ of the track or hit tower closest to $\eta=0$ in the (anti)proton
direction at $\protect\sqrt s=1800$ GeV: 
(a) data; (b, c, d) MC generated non-diffractive (ND), single-
(SD) and double-diffractive (DD) events.
The MC distributions are normalized by a fit to the data described in the text.}
\end{figure}

\newpage
\begin{figure}
\centerline{\psfig{figure=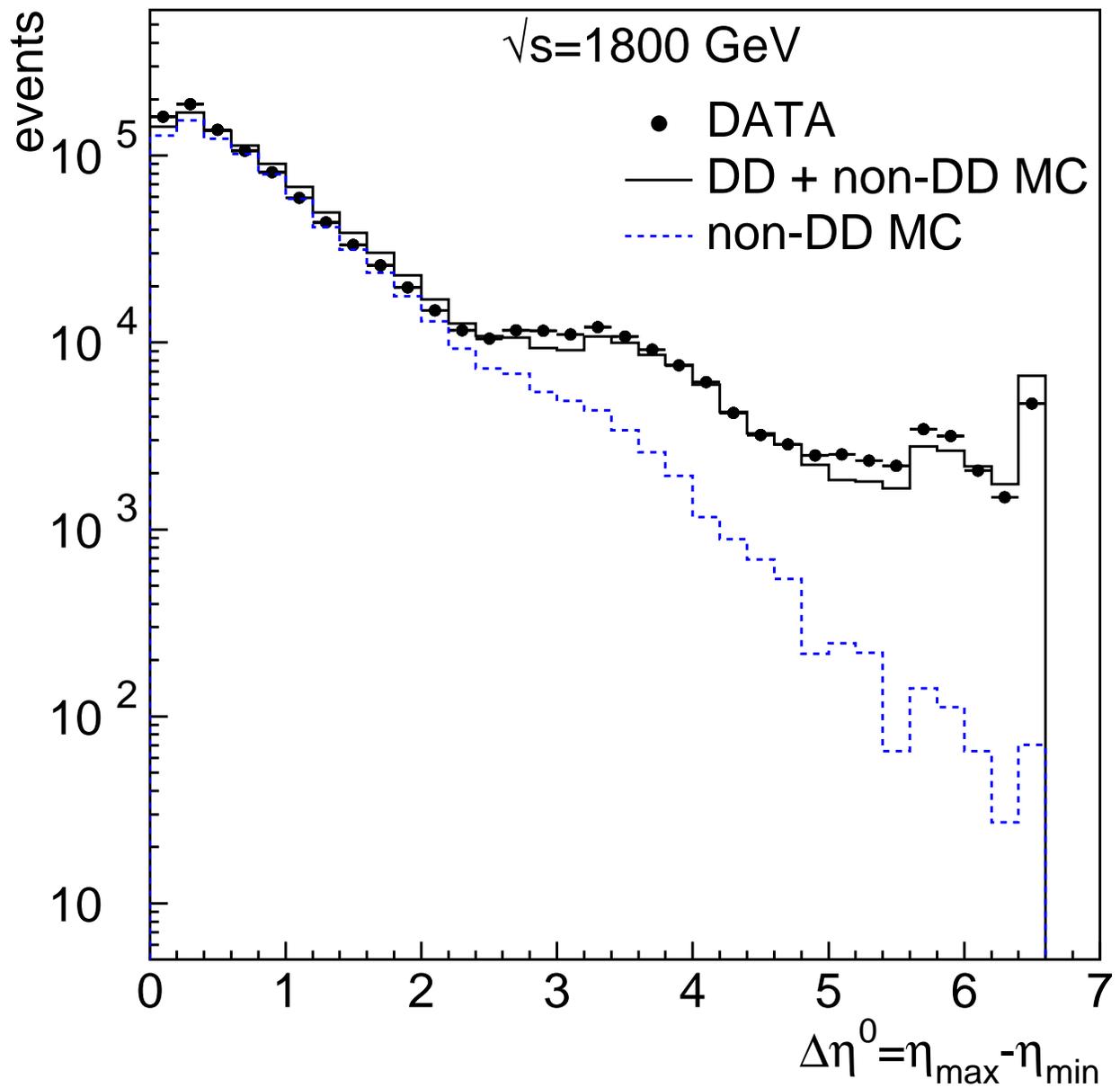,width=0.98\linewidth}}
\vfill
\caption{The number of events as a function of
$\Delta\eta_{exp}^0=\eta_{max}-\eta_{min}$ for data at 
$\protect\sqrt s=1800$ GeV (points), for double diffractive (DD) plus 
non-DD (MC) generated events (solid line), and for only non-DD
MC events (dashed line). The non-DD events are a mixture of 
97.3\% non-diffractive and 2.7\% single diffractive.
}
\end{figure}
\newpage
\begin{figure}
\centerline{\psfig{figure=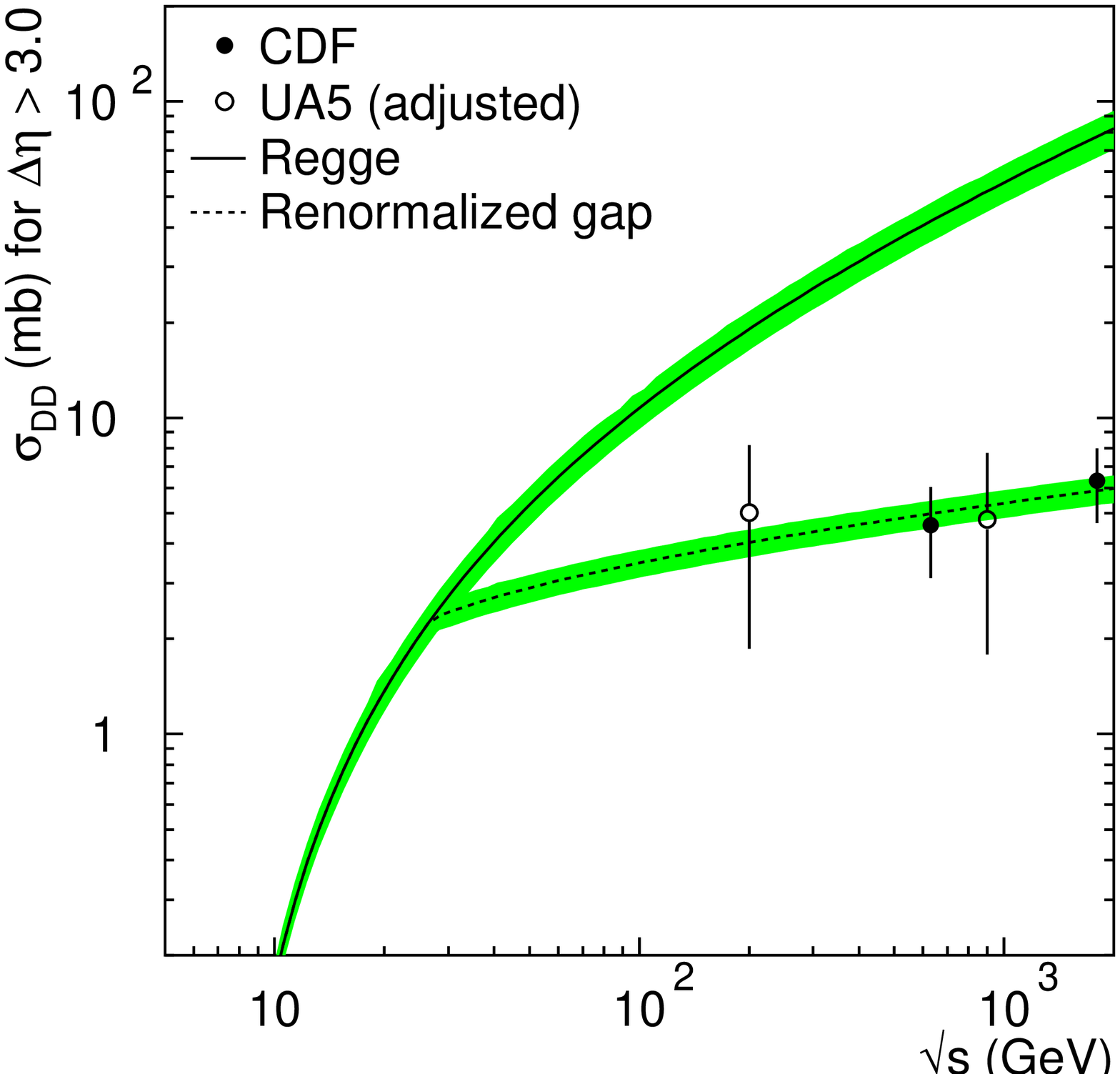,width=0.98\linewidth}}
\vfill
\caption{The total double diffractive cross section for 
$p(\bar p)+p\rightarrow X_1+X_2$ versus $\protect\sqrt s$ compared 
with predictions 
from Regge theory based on the triple-Pomeron amplitude and 
factorization (solid curve) and from the renormalized gap probability model 
(dashed curve).}
\end{figure}


\begin{references}
\bibitem{pseudo}  We use rapidity and pseudorapidity, $\eta$, interchangeably;
$\eta\equiv -\ln(\mbox{tan}\frac{\theta}{2})$,
where $\theta$ is the polar angle of a particle with respect
to the proton beam direction. The azimuthal angle is denoted by $\phi$, 
and transverse energy is defined as $E_T=E\sin\theta$.
\end{references}
\end{document}